\documentclass[fleqn,usenatbib]{mnras}

\usepackage[T1]{fontenc}

\DeclareRobustCommand{\VAN}[3]{#2}
\let\VANthebibliography\thebibliography
\def\thebibliography{\DeclareRobustCommand{\VAN}[3]{##3}\VANthebibliography}

\usepackage{graphicx}	
\usepackage{amsmath}	
\usepackage{amssymb}	
\usepackage{xcolor}

\newcommand{\cyan}[1]{\textcolor{cyan}{#1}}

\title[OJ~287 Blazar]{Spectral States of OJ~287 blazar from Multi-wavelength Observations with $AstroSat$}
\author[K. P. Singh et al.]{K. P. Singh,$^{1,7}$\thanks{E-mail: kpsinghx52@gmail.com (KPS)}
P. Kushwaha,$^{2}$\thanks{pankaj.kushwaha@aries.res.in (PK)}\thanks{Aryabhatta Postdoctoral Fellow}
A. Sinha,$^{3,6}$
Main Pal,$^{4, 8}$
A. Agarwal, $^{5}$
\newauthor
G. C. Dewangan$^{6}$
\\
$^{1}$Department of Physical Sciences, Indian Institute of Science Education and Research Mohali, Knowledge City, Sector 81, SAS Nagar, Punjab 140306, India\\
$^{2}$ Aryabhatta Research Institute of Observational Sciences (ARIES), Manora Peak, Nainital 263001, India\\
$^{3}$ Laboratoire Univers et Particules de Montpellier, Université de Montpellier, CNRS/IN2P3, CC 72, Place Eugène Bataillon, F-34095 Montpellier Cedex 5, France\\
$^{4}$ Centre for Theoretical Physics, Jamia Millia Islamia, New Delhi-110025, India \\
$^{5}$ Raman Research Institute, C. V. Raman Avenue, Sadashivanagar, Bengaluru 560080, India\\
$^{6}$Inter-University Centre for Astronomy and Astrophysics, Ganeshkhind, Pune 411 007, India \\
$^{7}$Department of Astronomy and Astrophysics, Tata Institute of Fundamental Research, 1 Homi Bhabha Road, Mumbai 400005, India\\
$^{8}$Department of Physics, Ramjas College, University of Delhi, New Delhi-110007, India\\
}

\date{Accepted XXX. Received YYY; in original form ZZZ}

\pubyear{2015}

\begin{document}
\label{firstpage}
\pagerange{\pageref{firstpage}--\pageref{lastpage}}
\maketitle

\begin{abstract}
We present {\it AstroSat} soft X-ray, near-UV (NUV) and far-UV (FUV) observations of a blazar, OJ~287, carried out in 2017, 2018, and 2020. The  simultaneous observations with NuSTAR in 2017 provide a broad-band look encompassing NUV, FUV, soft and hard X-rays. Captured in three different broadband spectral states in three observations, the X-ray spectrum is found to be the hardest during 2018, while the high-energy-end of the simultaneous optical-FUV spectrum shows a steepening that is modeled with a broken power-law spectrum. The spectral energy distribution (SED) in 2017  shows a relatively flatter optical-FUV and soft X-ray spectra, implying an additional emission component. The 2020 optical-FUV spectrum is harder than in 2017 and 2018, with an extremely soft X-ray spectrum and a hardening above $\sim$1 GeV, similar to the SEDs of High-energy-peaked BL Lac objects (HBL), thereby establishing that this additional emission component has HBL-like properties. The {\it AstroSat} multi-wavelength observations trace the spectral evolution from the end-phase of the HBL component in 2017 to its disappearance in 2018 followed by its revival in 2020. A single zone leptonic model reproduces the 2018 broadband spectrum while the 2017 and 2020 SEDs require an additional HBL-like emitting zone. The spectral evolution of the high-energy-end of optical-UV spectrum, revealed by the FUV observations in 2017 and 2018, strongly suggests that X-ray spectral changes in the normal broadband spectral state of OJ~287 are primarily due to the evolution of the optical-UV synchrotron spectrum.
\end{abstract}

\begin{keywords}
BL Lac objects: individual: OJ 287 -- galaxies: active -- galaxies: jets --
radiation mechanisms: non-thermal -- gamma-rays: galaxies -- X-rays: galaxies
\end{keywords}



\section{Introduction}
BL~Lac type objects are a special class of active galactic nuclei known as blazars, that are powered by accretion on to a super-massive black-hole (SMBH) and  have a super-luminal relativistic jet directed almost along the line-of-sight to us. The radiation from such objects covers the entire range of electromagnetic spectrum from radio to very high energy (VHE, E $>$ 100 GeV) $\gamma$-rays and can be highly polarised.  The entire emission originates almost exclusively within the jet and exhibits a broad spectral energy distribution (SED) with two distinct humps ascribed to synchrotron and inverse-Compton processes, and with their peaks shifting towards  lower frequencies with increasing bolometric luminosity thus giving rise to a spectral sequence that has been clubbed into
three classes known as Low-frequency-peak BL Lac (LBLs), intermediate-frequency-peak BL
Lacs (IBLs), and high frequency peak BL Lacs (HBLs) \citep{Fo1998,Abdo10}. X-ray emission from blazars is either due to Synchrotron (Sy) process from highly energetic electrons (or protons) or due to Synchrotron Self-Compton (SSC) process where low frequency photons are boosted in energy by the inverse Compton (IC) process from the same electrons that produced the Sy emission. 
Radiation from these objects is highly variable in intensity and spectral shapes on time-scales of minutes to years, indicative of highly compact yet powerful sub-parsec size for the  emission regions powering these objects.

 OJ~287 is an archetypal BL Lac object with red-shift, z, of 0.306 \citep{Sit85,Nil10}, belonging to the class of LBLs.  Ever since its discovery and identification of its optical counterpart, OJ~287 has been found to be a very dynamic source \citep{Wenz71,Sit85}. These studies have unearthed some peculiar features, uncommon for blazars, like a peculiar periodic outbursts at an interval of $\sim$12 yr \citep{Sill96a}, suggesting the possible presence of a binary SMBH system in its nucleus.
Other prominent features are an apparent periodicity of a few years in the position of quasi-stationary radio knots seen at 15 GHz, as well as a $\sim$22~yr periodicity \citep{Br18}, and a claim of $\sim$30-yr period from another study \citep{Co17}. These inferences from radio bands
are the basis of claim of a precessing and rotating jet in OJ 287, and models based on precession for the recurring $\sim$12-yr optical flares \citep{Co17,Br18,Butu20}.

OJ~287 has been monitored regularly over the years with the {\it Swift} X-ray Telescope (XRT).
It has been very active since mid-November 2015, in the optical wavebands \citep[e.g.][]{Valt16,Gup17, Gup19} to X-ray energies \citep{Komo17, Grupe17,Kush18a, Kush18b, Kap18,Komo20}. These
activities have been concurrent with the predicted (and observed) secondary BH impact-induced flares of 2015 and 2019  in the optical and infra-red wavelengths \citep{Valt10,Valt16,Dey18,Laine20},
indicating a connection between the two \citep{Komo20, Kush20a}. In this scenario, X-ray activities are also expected by the impacts of the secondary BH in the accretion disk leading to enhanced accretion and jet activity of the primary black hole. This binary BH (BBH) model 
can also reproduce the observed temporal variations in the position angle of OJ~287's radio jet \citep{Dey21}.

Multi-wavelength (MW) studies of activities since 2015 outburst show strong variations in the broad-band spectra of the source \citep{Komo17, Kush18a, Kush18b, Kap18, Main20, Komo20,Kush20b}.
Contrary to a simple power-law that describes the spectrum in different energy bands quite well, drastic spectral 
changes have been reported at X-ray energies \citep[e.g.][]{Komo17,Kush18b,Main20,Komo20,Kush20a}. A soft X-ray excess\footnote{This is different from the soft X-ray excess claim of \citet{Komo21a} where they call the extremely soft X-ray spectral state as the soft X-ray excess.} above the power-law, usually seen in Seyfert galaxies, has also been reported in OJ~287 a few months before the expected disk-impact activity of December 2015 from observations with {\it XMM-Newton} \citep{Main20}. The spectrum, however, hardened during the 2015 impact outburst \citep[MJD: 57361;][]{Kush18a} consistent with a power-law photon spectral index, $\Gamma$= 1.5 - 1.6. 
Broadband SEDs of 2015 activity revealed a new spectral state of the source with significant spectral evolution in all the bands from NIR-optical to MeV-GeV gamma-ray \citep[e.g.][]{Gup17,Kush18a,Kap18,Kush20a}. It revealed a
sharp break in NIR-optical spectrum deviating strongly from its well-known and established 
power-law form consistent with a thermal emission component \citep{Kush18a,Rod20}.
The first appearance of this departure was traced back to May 2013 \citep[MJD 56439;][]{Kush18a,Kush20a}.
A concurrent hardening of MeV-GeV spectrum as well as a shift in the location of the peak of the high-energy
hump was also seen which is consistent with both leptonic and hadronic emission scenarios \citep{Kush18a,Rod20}.
Interestingly, with the end of 2015 MW activity, the NIR-optical spectral break disappeared as well \citep{Kush20a}.
A similar NIR-optical spectral break has been seen during 2020 activity of the source as well \citep{Kush20b}.
 The spectral and temporal coincidence of the sharp NIR-optical spectral break seen in 2015
with the elements of the BBH model added an independent evidence in favor of the BBH model \citep{Kush20b}. However, the timing of its re-appearance in 2020 activity indicates some missing physics \citep{Kush20b}.

The most surprising and unexpected trend was the strongest UV to X-ray activity that started \(\ 7-8\) months after the disk-impact flare of December 2015 \citep{Komo17,Komo20,Komo21b}. Detailed MW studies showed that this activity was driven by major changes
in the X-ray spectral state of the source \citep{Komo17,Kush18b,Komo20,Komo21a} with the SED
of this episode being similar to the high-frequency-peaked (HBL) like emission component of blazars, having the peak of the low-energy hump in UV to soft-X-ray region in the SED plot
\citep{Kush18b,Kush20b,Prince2021a,Prince2021b}. A few months after the start of this activity, the source was detected for the first time at very high energies (VHEs) by the VERITAS facility and it remained active at VHE for about six months \citep{Brien17}. OJ~287 underwent another similar high optical to X-ray flux in 2020, with the X-ray flux (0.3-10 keV) reaching the second highest reported for this source \citep{Komo20,Komo21a}. 
Spectral studies confirm this to be the re-emergence of the soft
X-ray spectral state \citep[and references therein]{Komo17,Grupe17,Komo20} with broadband SED similar to the blazars HBL component \citep{Kush20a,Prince2021b}.
The spectral similarity of the broadband emission to the blazars HBL component \citep{Kush18b,Kush20b} as well
as the temporal and polarization \citep[and references therein]{Komo20} behavior points
to a jet origin. However, the activity driven by HBL-like component has shown some peculiar flux  variability trends \citep{Kush18b,Komo20,Kush20a} that seem characteristic of OJ~287 and
contrary to the normal behavior known for this source. Another characteristic feature
linked to this state is the NIR-optical spectral break that is rather sharp and inconsistent
with the broad-band jet origin emission \citep{Kush20a,Kush20b,Rod20,Kush18a}. Also, for the first time, an iron line absorption feature was reported for this source during its 2020
high activity \citep{Komo20}. 
 
Here, we report spectral and temporal observations of OJ~287 with the {\it AstroSat} \citep{Si2014} -- a multi-wave Indian Space Observatory.  The observations were performed during 2017, 2018, and 2020.
A detailed spectral and temporal study of multi-wave data from these three observations is presented here. We include concurrent data from a few other facilities. 
We also present our analysis of the {\it NuSTAR} observation carried out simultaneously with the {\it AstroSat} observations in 2017.  
The next section gives the details of data, followed by data reduction procedures in \S3. In \S4 we present the spectral analysis of X-ray data and results obtained, followed by modelling of multi-wave SED in \S5 and a discussion in \S6. Our summary and conclusions are in \S7.

\section{Observations}
\begin{table*}
	\centering
	\caption{Log of Observations of OJ~287. UVIT Filters used are: NUV F3 = F245M; NUV F5 = F263M; FUV F1 = F148W; FUV F2 = F154W; FUV F3= F169M; FUV F5=F172M; and FUV F7 = F148Wa, where the numbers 245, 263, 148, 154, 169 and 172 correspond to the central wavelength of the corresponding filter in nm. More information on filters is available at: https://uvit.iiap.res.in/Instrument/Filters}
	\label{tab:table1}
	\resizebox{2\columnwidth}{!}{
	\begin{tabular}{lcccll} 
		\hline
	Instrument & Observation ID & Start Time (UT) & Stop Time (UT) & Exposure (s) & Count Rate\\
                     &  & Y:M:D:H:M:S & Y:M:D:H:M:S &  & SXT; FPMA,B; NUV/FUV \\
                        & &  & & & 0.3-7.0 keV; 3-30 keV; F3/F5 \\
		\hline	
        NuStar FPMA & 90201054002 & 2017:04:09:11:09:31 & 2017:04:10:15:27:31 & 51580 & 0.0823$\pm$0.0013\\
	NuStar FPMB & 90201054002 & 2017:04:09:11:09:31 & 2017:04:10:15:27:31 & 51680 & 0.0815$\pm$0.0013\\
	AstroSat SXT & 9000001152 & 2017:04:09:22:05:31 & 2017:04:11:00:24:00 & 21930 & 0.152$\pm$0.0031\\
	AstroSat SXT & 9000002040 & 2018:04:15:10:58:33 & 2018:04:21:06:57:27 & 110500 & 0.066$\pm$0.0012\\
	AstroSat SXT & 9000003672 & 2020:05:15:09:22:37 & 2020:05:20:03:05:58 & 60410 & 0.232$\pm$0.0022\\
    AstroSat NUV F3 & 9000001152 & 2017:04:09:22:10:47 & 2017:04:10:09:30:47 & 4050  &  7.45$\pm$0.40\\
    AstroSat NUV F5 & 9000001152 & 2017:04:10:09:42:12 & 2017:04:10:15:52:12 & 3400   & 6.38$\pm$0.37 \\
    AstroSat FUV F3 & 9000001152 & 2017:04:09:22:06:41 & 2017:04:10:15:54:41 &   8000 &  1.87 $\pm$0.20\\
	AstroSat FUV F3 & 9000002040 & 2018:08:15:06:03:50 & 2018:08:16:02:03:42 & 25150 &  1.05$\pm$0.15\\
	AstroSat FUV F5 & 9000002040 & 2018:08:15:06:03:50 & 2018:08:16:02:03:42 &  29400 &  0.51$\pm$0.11\\
	AstroSat FUV F1 & 9000003672 & 2020:05:15:15:52:13 & 2020:05:15:17:35:31 & 4924 &  6.53$\pm$0.24\\
	AstroSat FUV F2 & 9000003672 & 2020:05:15:17:37:30 & 2020:05:16:09:46:42 &  4198 &  5.40$\pm$0.23\\
	AstroSat FUV F5 & 9000003672 & 2020:05:15:06:03:50 & 2020:05:16:02:03:42 & 8992 &  2.16$\pm$0.14\\
	AstroSat FUV F7 & 9000003672 & 2020:05:16:09:48:41 & 2020:05:17:13:27:56 &  5475 &  6.51$\pm$0.23\\
		\hline
	\end{tabular}}
\end{table*}

{\it AstroSat}  carried out three extended observations of OJ~287 in 2017, 2018 and 2020 with the Soft X-ray Telescope (SXT) covering the energy band of 0.3$-$7.1 keV \citep{Si2016,Si2017}, and with the Ultra-Violet Imaging Telescope (UVIT) simultaneously covering the NUV and FUV bands\citep{Tan17, Tan20}. In each observation, OJ~287 was observed with the SXT throughout an orbit of the satellite taking care that the Sun avoidance angle is $\ge$ 45$^\circ$ and RAM angle (the angle between the payload axis to the velocity vector direction of the spacecraft) > 12$^\circ$ to ensure the safety of the mirrors and detector. 
The UVIT consists of two 35cm Ritchey-Chrètien telescopes with one telescope dedicated to the far ultraviolet (FUV: 1250 – 1830 Å). The second telescope uses a dichroic beam splitter that splits the beam into the near ultraviolet (NUV: 1900 – 3040 Å) and the visible (VIS: 3040 – 5500 Å) channels.  Each waveband has a choice of filters as given in \citep{Tan17, Tan20}. The VIS channel is used for spacecraft tracking and data from it are not usable for science. 
OJ~287 was monitored in the NUV and FUV, simultaneous with X-ray instruments, in all the observations. The 2017 observations were also simultaneous with the {\it NuSTAR} \citep{Har13} which detects hard X-rays in the energy range of 3$-$78 keV. {\it NuSTAR} has two co-aligned hard X-ray telescopes each equipped with detectors 
in their focal plane and known as Focal Plane Module A and B (FPMA, FPMB) respectively.  All the observations, their start and stop times, useful exposure times after screening (see below) and observed count rates are listed in Table~\ref{tab:table1} for all the instruments from which data were used in this paper. The filters used for the FUV and NUV observations are also given in Table~\ref{tab:table1}. 
OJ~287 was also observed simultaneously with the Large Area Xenon Proportional Counter \citep{Ant17} aboard {\it AstroSat}, however, no useful data were obtained from this instrument.

\section{Data Reduction}

{\bf AstroSat-SXT:} Data from individual orbits (Level~1) were received at the SXT POC (Payload Operation Centre) from the ISSDC (Indian Space Science Data Center). The Level~1 data were processed using SXTPIPELINE software at the POC where events were selected with event grading similar to {\it Swift}-XRT, 
and events with grades > 12 were removed. All events above a preset threshold were time-tagged, applied with the coordinate transformation from raw (detector) to sky coordinates, and bias-subtracted. The flagging of bad pixels, the PHA construction 
for each event, the conversion from the event PHA to PI and a screening for hot and flickering pixels were also carried out in the pipeline. 
The events were further screened for bright Earth avoidance angle of $\ge$ 110$^{\circ}$. Data taken during the passage through the South Atlantic Anomaly (SAA) were also removed based on the criterion that the Charged Particle Monitor (CPM) rate is below 12 counts s$^{-1}$. 
Good Time Intervals (GTI) were generated after all the screening and calibrations mentioned above were applied and Level~2 data events files were produced. All the GTI's during each orbit were selected, time overlaps between consecutive orbit data files removed and a merged events file of all cleaned events was generated using a Julia tool made by the SXT team. X-ray light curves and spectra were extracted from data merged from all the orbits of the {\it AstroSat}. Source counts were extracted from a circular region of 12\arcmin radius for both the light curves and spectra \citep{Si2016,Si2017}.

{\bf NuSTAR:} Data from {\it NuSTAR} \citep{Har13} observations were processed with the NuSTARDAS software package v1.8.0 available within HEASOFT package (6.28).  The calibration database CALDB (v. 20180126 update) was used for analysis. After running {\it nupipeline} on each observation, {\it nuproducts} was used to obtain the light curves and spectra. A circular region of 44\arcsec radius centered on OJ~287 was used for extracting the source counts. Background was obtained from a circular region of 100\arcsec radius on the same detector chip located 200\arcsec away from OJ 287. The count rate observed from OJ~287 in either of the detectors was 0.08 counts s$^{-1}$. The spectra were grouped to ensure a minimum of 50 counts per bin for the spectral analysis.

{\bf AstroSat-UVIT:} The Level-2 UVIT data consisting of photon lists and images were obtained from the ISSDC.  These data were processed with the UVIT pipeline version 6.3 and the CALDB version 20190625. 
We used the {\it curvit}\footnote{\url{https://github.com/prajwel/curvit}} package \citep{Joseph21} to extract the light curves in each filter with a time-bin of 240 seconds.  A circular region with a radius of 2.5\arcsec (plate scale of 0.416\arcsec per pixel) was used as the source region while a nearby circular region of 5.0\arcsec radius was used for the background. The same procedure was used for each filter.
Finally, the measured count-rate was converted into flux using the zero-point of the respective UVIT bands mentioned in the \citet{Tan20}. The flux values were then corrected for reddening using E(B-V) = 0.0241 with reddening coefficients estimated following \citet{Cardelli89}.

\section{Analysis and Results}

\begin{figure}
	\includegraphics[width=\columnwidth]{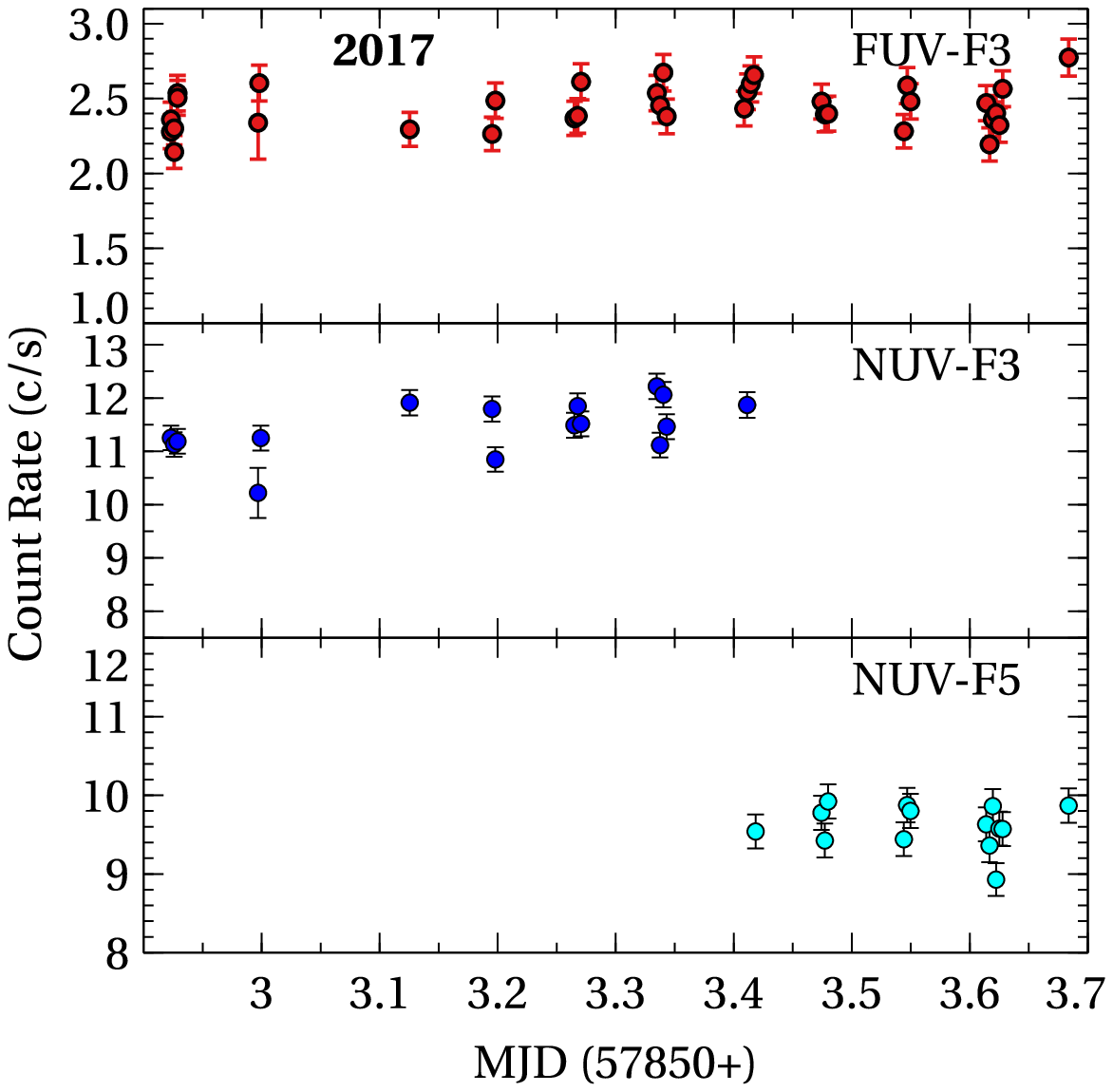}
	\includegraphics[width=\columnwidth]{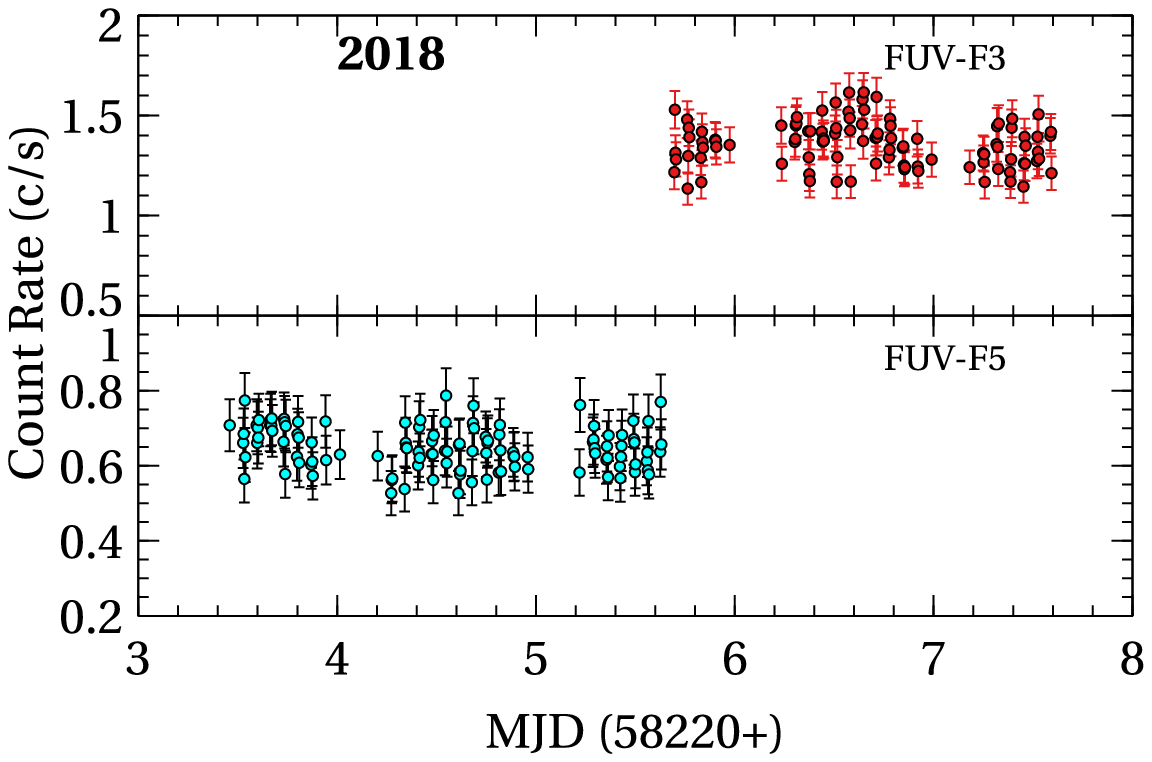}
	\includegraphics[width=\columnwidth]{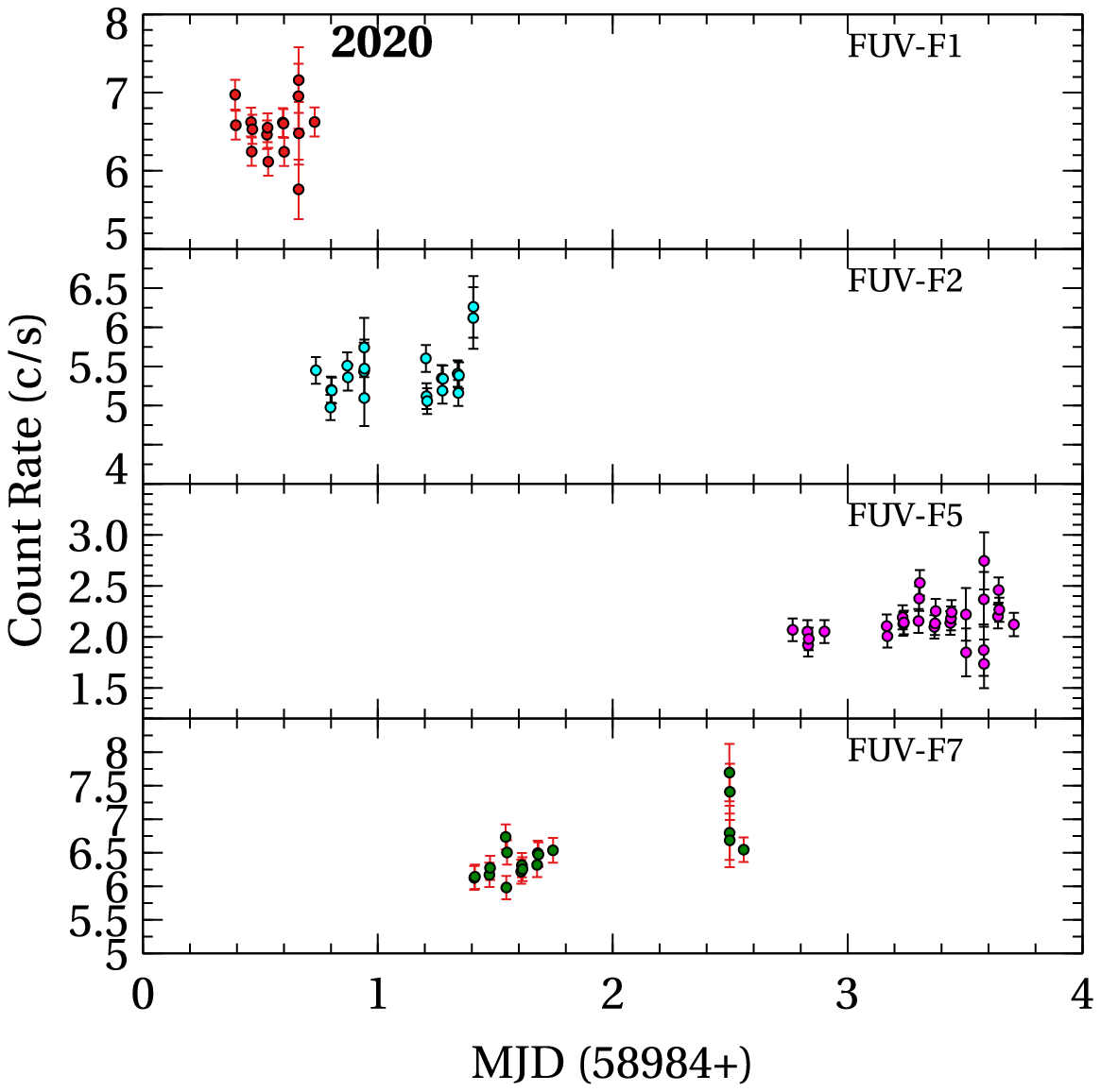}
    \caption{NUV and FUV light curves of OJ~287 constructed from the 2017, 2018, and 2020 observations performed with the {\it AstroSat}-UVIT}.
    \label{fig:uvit}
\end{figure}

\subsection{Near and Far UV Variability}\label{subsec:UVITlc}
Long simultaneous exposures with the {\it AstroSat} UVIT and SXT in 2017 and 2018 allow us to explore short-term UV variability during the X-ray observations. Figure \ref{fig:uvit} shows the light curve of the source extracted with a time-bin of 4-min in different UVIT bands. The uniformly looking gaps are due to spacecraft duty cycle for each orbit. Though  visually it appears that there is a variability on small time scales, to check whether its statistically significant or just due to random fluctuations, we estimated the excess variance given by \citep{vaughan2003}
\begin{align*}
F_{var} &= \sqrt{\frac{S^2 - \overline{\sigma_{err}^2}} {{\bar{x}}^2}} \\
err(F_{var}) &= \sqrt{\left(\sqrt{\frac{1}{2N}} \frac{\overline{\sigma_{err}^2}}{{\bar{x}}^2 F_{var}}\right)^2 + \left(\sqrt{\frac{\overline{\sigma_{err}^2}}{N}} \frac{1}{\bar{x}})\right)^2}
\end{align*}

for each of the UVIT filter bands  where $\rm S^2, \bar{x},$ and $\overline{\sigma_{err}^2}$
are respectively the
variance, mean, and mean of the squared-error of the count-rate and N is the number of the data points. The results indicate that the source is variable at both NUV (\(\rm F_{var}^{F3}=0.038\pm0.006; F_{var}^{F5}=0.017\pm0.008\)) and FUV (\(\rm F_{var}^{F3}=0.03\pm0.01\)) energies during 2017 but only FUV-F3 (\(\rm F_{var}^{F3}=0.051\pm0.009\)) is variable during 2018 observations while all the FUV bands (\(\rm F_{var}^{F1}=0.03\pm0.01; F_{var}^{F2}=0.04\pm0.01; F_{var}^{F5}=0.06\pm0.02; F_{var}^{F7}=0.05\pm0.01;\)) show variability during the 2020 observation. The variability inferred in the NUV and FUV intensity, based on the formal variance test is to be taken very conservatively as the data and the timescales do not allow us to compare with other stars in the same field of view due to their faintness.
The instrumental effects, therefore, cannot be ruled out.  The results presented here are not affected by the presence or absence of this small time scale variability.

\subsection{X-ray Spectral Analysis and Results}

\begin{figure}
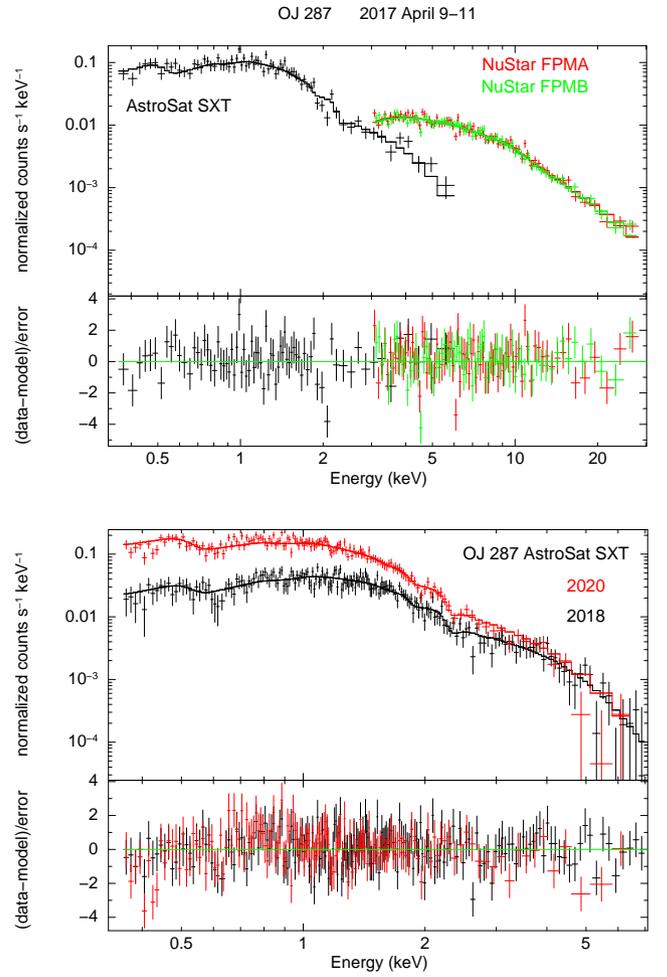

	\includegraphics[width=0.75\columnwidth, angle=270]{C_fixNh_zpow_gr50.eps}
	\vskip 0.5cm
	\includegraphics[width=0.68\columnwidth, angle=270]{2018_20_zpow_fixedNh.eps}
    \caption{{\bf Top}: X-ray spectra of OJ~287 obtained from simultaneous observations with the {\it AstroSat} SXT and the {\it NuSTAR} FPMA and FPMB in 2017.
    {\bf Bottom}: X-ray spectra of OJ~287 obtained from observations with the {\it AstroSat} SXT in 2018 and 2020. In both the plots, the histograms show the best fit models of a Galactic absorber and a red-shifted power-law.  The lower panels in both Top and Bottom figures show the residuals from the best fit models.}
    \label{fig:sxt171820}
\end{figure}

 The X-ray light curves were made in the energy band of 0.3-7.1 keV.  Light curves in the energy band of 7.1-10 keV were examined for any non-X-ray flares during the observations, and the GTIs corresponding to such events were considered as spurious and the data corresponding to those times were removed for any further analysis. The useful exposure times for the clean events
 are listed in Table~\ref{tab:table1}.  The X-ray light curves thus obtained showed that OJ~287 was in a steady/non-variable state during each of the three observations. Therefore, spectra obtained from all the merged events in each observation were used for spectral analysis.
A background spectral file ”SkyBkg\string_comb\string_EL3p5\string_Cl\string_Rd16p0\string_v01.pha”, derived from a composite of several deep blank sky observations, distributed by the instrument team was used for spectral analysis for all the spectra analysed here. We used the standard ancillary response file
(ARF) "sxt\string_pc\string_excl00\string_v04\string_20190608.arf". 
The spectral response file used in this work is sxt\string_pc\string_mat\string_g0to12.rmf. The response files and the background file are available at the SXT POC website\footnote{\url{https://www.tifr.res.in/~astrosat\_sxt/index.html}}.
The source spectra were grouped using the {\it grppha} tool to ensure a minimum of 50 counts per bin, prior to further analysis here and below. The observed count rates are given in Table~\ref{tab:table1}.

\begin{table*}[!ht]
	\centering
	\caption{Best fit spectral parameters for the two models M1: C*tbabs(zpowerlaw) and M2: C*tbabs(zlogpar), where two C values are the relative normalization values required for FPMA and FPMB respectively with respect to the SXT, used for fitting the X-ray spectra of OJ~287.  The errors quoted here are with 90\% confidence based on minimum $\chi^2$+2.71.}
	\label{tab:table2}
	\resizebox{2\columnwidth}{!}{
	\begin{tabular}{lcccccccccccc} 
		\hline
		Date & Detector & Model & C & N$_{H}$ & $\Gamma$ (PL) & $\alpha$ (logpar) & $\beta$ (logpar) & $\chi^2_{\nu}$/dof & Fx${_1}$ & Fx${_2}$ & Fx${_3}$ \\
		& & & & 10$^{20}$cm$^{-2}$ & & &    &   & 0.3$-$2.0 keV & 2$-$ 10 keV & 10$-$30 keV\\
		& & & & &  & &   &   & ergs cm$^{-2}$ s$^{-1}$ & ergs cm$^{-2}$ s$^{-1}$ & ergs cm$^{-2}$ s$^{-1}$ &\\
		\hline
		2017 & SXT+FPMA+B & M1 & 0.97$^{+0.07}_{-0.07}$,1.02$^{+0.07}_{-0.08}$ & 1.7$^{+1.9}_{-1.7}$ & 2.06$\pm$0.04 & - & - & 1.1244/221 & 5.9$\times$10$^{-12}$ &  5.0$\times$10$^{-12}$ & 3.1$\times$10$^{-12}$ \\
		
		 & SXT+FPMA+B & M1 & 0.97$^{+0.07}_{-0.07}$,1.01$^{+0.08}_{-0.07}$ & 2.4(fixed) & 2.07$\pm$0.03 & - & - & 1.1209/222 & 5.8$\times$10$^{-12}$ &  5.0$\times$10$^{-12}$ & 3.0$\times$10$^{-12}$ \\
		
		& SXT+FPMA+B & M2 & 1.03$^{+0.12}_{-0.10}$,1.08$^{+0.12}_{-0.11}$ & 4.5$\pm$4.0 & & 2.2$\pm$0.2 & -0.08$\pm$0.10 & 1.1216/220 & 5.8$\times$10$^{-12}$ & 4.8$\times$10$^{-12}$ & 3.1$\times$10$^{-12}$\\
		
		& SXT+FPMA+B & M2 & 1.00$^{+0.09}_{-0.08}$,1.04$^{+0.09}_{-0.08}$ & 2.4(fixed) & & 2.12$\pm$0.08 & -0.03$\pm$0.05 & 1.1201/221 & 5.8$\times$10$^{-12}$ & 4.8$\times$10$^{-12}$ & 3.1$\times$10$^{-12}$\\
		
	    2018 & SXT & M1 & - & 4.9$^{+3.0}_{-3.0}$ & 1.88$\pm$0.11 &  &  & 0.893/195 & 2.2$\times$10$^{-12}$ & 2.9$\times$10$^{-12}$ &  $-$\\
	    
	     & SXT & M1 & - & 2.4(fixed) & 1.80$\pm$0.06 &  &  & 0.897/196 & 2.3$\times$10$^{-12}$ & 3.1$\times$10$^{-12}$ &  $-$\\
	     
		 & SXT & M2 &  & 9.7$^{+10.0}_{-9.3}$ & $-$ & 2.2$^{+0.60}_{-0.57}$ & -0.31$^{+0.59}_{-0.57}$ & 0.894/194 & 2.2$\times$10$^{-12}$ & 3.2$\times$10$^{-12}$ &  $-$\\
		 
		 & SXT & M2 & - & 2.4(fixed)  & $-$ & 1.74$^{+0.11}_{-0.12}$ & 0.11$^{+0.20}_{-0.19}$ & 0.897/195 & 2.3$\times$10$^{-12}$ & 2.9$\times$10$^{-12}$ &  $-$\\
		 
	     2020 & SXT & M1 & - & 9.0$^{+2.2}_{-2.1}$ & 2.81$\pm$0.10 &  &  & 0.97/147 & 9.0$\times$10$^{-12}$ & 3.0$\times$10$^{-12}$ & $-$ \\
	      
	      & SXT & M1 & -  & 2.4(fixed) & 2.52$\pm$0.04 &  &  & 1.16/148 & 9.8$\times$10$^{-12}$ & 3.9$\times$10$^{-12}$ & $-$ \\
	      
		 & SXT &  M2 & - & 12.4$^{+7.0}_{-6.0}$ &  & 3.04$^{+0.45}_{-0.43}$  & -0.27$\pm$0.43 & 0.972/146  & 8.8$\times$10$^{-12}$ & 3.3$\times$10$^{-12}$ & $-$\\
		 
		 & SXT &  M2 & - & 2.4(fixed) &  & 2.4$^{+0.06}_{-0.07}$  & 0.47$\pm$0.17 & 1.012/147  & 9.4$\times$10$^{-12}$ & 2.8$\times$10$^{-12}$ & $-$\\
\hline
\end{tabular}}
\end{table*}

Spectra were modelled using the  spectral analysis package {\sc xspec} (version 12.9.1; \citealt{Ar1996}) distributed with the {\sc heasoft} package (version 6.20). We have tried two types of models to fit the observed spectra, viz., constant*Tbabs*zpowerlaw and constant*Tbabs*zlogpar, using the $\chi^2$ minimisation. Here, $Tbabs$, is a multiplicative model with the model parameter $N_\mathrm{H}$ , i.e., the equivalent neutral hydrogen column density along the line of sight. The elemental abundance table was set to $"aspl"$ \citep{As2009}. The models zpowerlaw and zlogpar are the redshifted power-law and the redshifted logparabola \citep{Mass2004} frequently used quite successfully in the X-ray spectral studies of blazars.  The redshift was fixed at 0.306 in both the models, while the other parameters like the $N_\mathrm{H}$, the photon index ($\Gamma$) in the case of a power-law were allowed to vary freely while minimising the $\chi^2$.  Similarly in the case of zlogpar model the parameters: $\alpha$ for the slope at the pivot energy of 1 keV (fixed) and $\beta$ for the curvature term were allowed to vary freely. We also fitted these models after fixing the $N_\mathrm{H}$ value to the Galactic value of 2.4$\times$10$^{20}$cm$^{-2}$ \citep[HI4PI Collaboration;][]{Bekhti2016}.

X-ray spectra of OJ~287 obtained from simultaneous observations with the {\it AstroSat} SXT and the {\it NuSTAR} FPMA and FPMB carried out in 2017 are shown in Figure~\ref{fig:sxt171820}. The results of the simultaneous spectral fitting to the three data sets are given in Table~\ref{tab:table2}. The inferred intensities based on the models used here are also listed in the Table~\ref{tab:table2} for three energy bands: 0.3-2.0 keV, 2.0-10.0 keV and 10-30 keV (above 30 keV, {\it NuSTAR} data is noise dominated\footnote{The hardening at high-energy-end of the X-ray spectrum for the 2017 observations
presented in \citet{KushICRC2021} is not real and is due to the noise.}). All the spectral parameters were tied together and varied jointly, except for the normalisation constants for the FPMA and FPMB detectors which were kept as variables while that for the SXT was kept frozen to 1.0.  The best fit parameters with their 90\% confidence errors based on minimum $\chi^2$+2.71 are listed Table~\ref{tab:table2}. The cross-normalization between detectors is found to be better than 10\% including statistical errors. An excellent fit ($\chi^2_{\nu}$=1.12) was obtained with the same values for all the spectral parameters for data from all the three detectors. The best fit N$_\mathrm{H}$ value of 1.7$^{+1.9}_{-1.7}$ $\times$ 10$^{20}$ cm$^{-2}$ is consistent with the Galactic value 2.4$\times$ 10$^{20}$ cm$^{-2}$, thereby implying no evidence for extra absorption towards OJ~287 for the best fit power-law model.  The same two models (constant*Tbabs*zpowerlaw and constant*Tbabs*zlogpar) were also fitted with fixed N$_\mathrm{H}$. The $\chi^2_{\nu}$ values for the best fit show that the two models are equally good fit to the spectra with or without fixed N$_\mathrm{H}$. 

The SXT spectra obtained for 2018 and 2020 observations were fitted to the same two models as above, individually.  All the parameters: column density, power-law index (or $\alpha$ and $\beta$) and the normalization were allowed to vary independently, and also with fixed column density as mentioned in \S4.  The results of X-ray spectral modelling are presented in Table~\ref{tab:table2}. The two models used here provide an equally good fit to the spectra presented in Figure~\ref{fig:sxt171820}.
The 2017 data, best fit model, constant*Tbabs*zpowerlaw, folded into the detector response and the residuals are shown in the top panel of
Figure~\ref{fig:sxt171820}.
Similarly, two X-ray spectra obtained from observations with the SXT in 2018 and 2020 and the folded models and residuals from the best fit Tbabs*zpowerlaw model are shown together in the bottom panel of Figure~\ref{fig:sxt171820}.

\begin{figure}
	\includegraphics[scale=0.5]{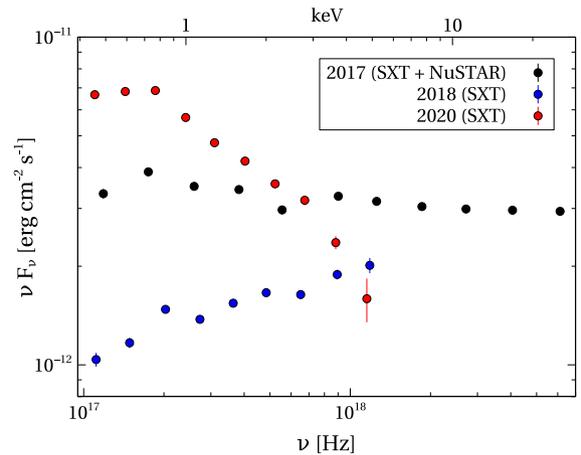}
    \caption{X-ray SEDs of OJ~287 from the three observations. All the SEDs
    correspond to the respective best-fit power-law model with N$_\mathrm{H}$ value fixed to the
    Galactic value.}
    \label{fig:xSED}
\end{figure}

 A considerable long-term variation over the years from 2017 to 2020 is observed in the X-ray spectra of OJ~287 \citep{Komo17,Komo20,Kush18b,Kush20a}. The X-ray flux decreased from 2017 to 2018 but increased in 2020.  The best fit simple power-law models show that $\Gamma$ is changing significantly across the years.  Equally good fits with the fixed N$_\mathrm{H}$ value in both the models show that there is no significant absorption in the source.
The best fit photon indices for a simple power-law show that the X-ray spectrum of OJ~287 steepens significantly with the intensity of the source in these three observations as is clear from the X-ray SEDs shown in Figure \ref{fig:xSED}.

The source is suspected to be variable in the NUV and FUV bands of UVIT during 2017 while it is variable only in FUV-F3 during 2018 (see Fig. \ref{fig:uvit}, Section 4.1). In 2020 data, all the FUV
bands show variability. Therefore, to study the possible implications of the variability on emission processes, we investigated and modeled the multi-band spectral energy distribution for observations as detailed in the next section.

\subsection{Fermi-LAT Spectral Analysis and Results}
We have used contemporaneous \textit{Fermi}-LAT \citep[Large Area Telescope;][]{lat09} PASS 8 (P8R3) instrument response function processed data to extract the MeV-GeV spectrum following the standard reduction procedure. Integrating the LAT data over the duration of {\it AstroSat} monitoring, the source is too weak
to reach a detection. We, therefore, integrated the LAT data over 2-3 months to obtain the average $\gamma$-ray spectrum of the source around the  {\it Astrosat} observations. Any shorter term variations in amplitude and spectrum are lost in the process.

The analysis was carried out following the standard recommended procedure for
the ``Binned Analysis for point sources\footnote{\url{https://fermi.gsfc.nasa.gov/ssc/data/analysis/scitools/binnededisp_tutorial.html}}'' using \textsc{Fermitools} (v2.0.8)
with \textsc{Fermipy} (v1.0.1). 
We selected only the events tagged
as ``SOURCE'' class (evclass=128, evtype=3) with energies > 80 MeV from a region of interest (ROI) of 15\textdegree centered on the source and a zenith angle cutoff of
90\textdegree. We then generated good time intervals using the recommended expression 
``(DATA\_QUAL>0)$\&\&$(LAT\_CONFIG==1)''. The spectral model for the ROI was generated
from the Fourth {\it Fermi}-LAT source catalog \citep[4FGL;][]
{4fgl}. The Galactic and isotropic diffuse contribution were accounted through the
template file gll\_iem\_v07.fits and iso\_P8R3\_SOURCE\_V3\_v1.txt provided by the
LAT team.

We performed a ``binned likelihood analysis'' with energy dispersion
taken into the account. During the likelihood fitting, the minimum fit energy range
was fixed to $\geq$ 100 MeV. The 80 MeV cut mentioned-above was to account for
the energy dispersion. Sources with Test Statistics (TS) $\rm \leq$ 0 were removed
while the spectral shape was fixed to the 4FGL values for source with TS $\leq$ 16. Using the best-fit model, we then extracted the SED in six energy-bins
(GeV): 0.1--0.3, 0.3--1.0, 1--3, 3--10, 10-100, 100--300, and 300 -- 800.

The overall source gamma-ray spectrum was modeled with a log-parabola model while the SED
was extracted using a power-law approximation of the best-fit spectrum in each
energy bin. The best-fit spectral parameters of log-parabola model for 2017, 2018, and 2020 LAT
spectrum are $\rm\alpha=2.08\pm0.25,~2.42\pm0.15,~1.72\pm0.23$, and $\rm\beta=0.07\pm0.11,~
-0.04\pm0.06, ~0.05\pm0.06$, respectively.  The resultant MeV-GeV spectra covering
the SXT pointings are shown in Figure \ref{fig:mwSED}.

\section{Broad-band SED and Modelling}\label{sec:model}

\begin{figure}
	\includegraphics[width=\columnwidth]{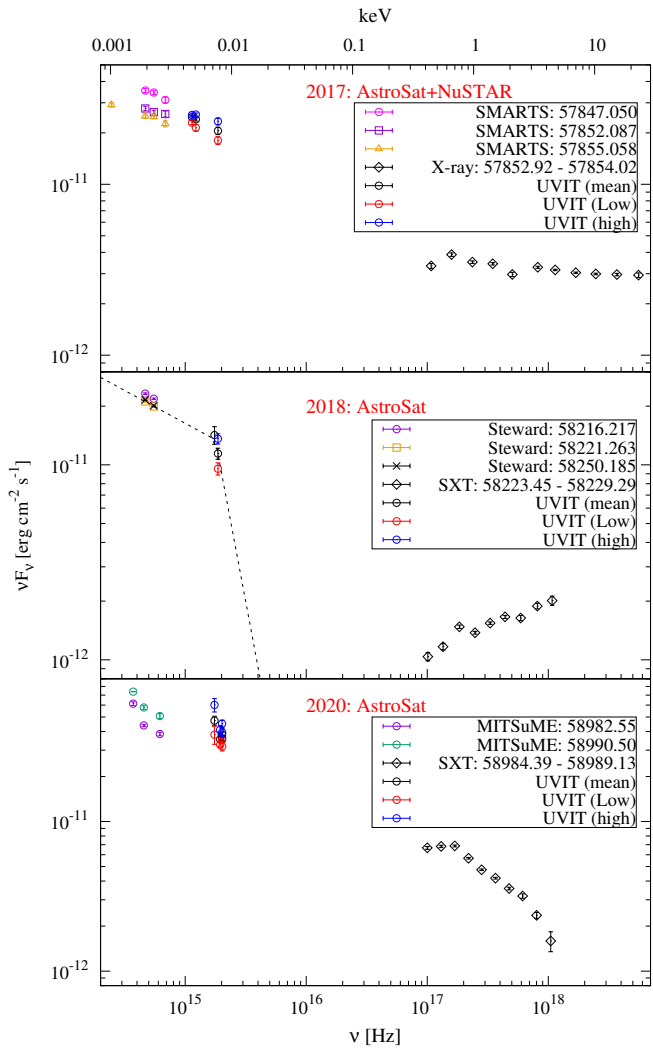}
    \caption{Optical to X-ray SEDs of OJ 287 from the three {\it AstroSat} observations along
    with contemporaneous NIR-optical data from public archives (see \S\ref{sec:model}). For the UVIT bands showing variability, we have plotted the highest, lowest and the mean fluxes in blue, red, and black respectively (see
    \S\ref{subsec:UVITlc}).}
    \label{fig:SED}
\end{figure}

Deep exposures with the SXT and the availability of simultaneous FUV data from the {\it AstroSat}, provide a direct view of yet unobserved high-energy-end of the optical-UV spectrum that has strong implications on the effect of synchrotron component
in driving X-ray spectral changes \citep{Isobe01,Kush13,Kush18b}.  
Additionally, the availability of {\it NuSTAR} data during 2017 allows us to constrain the IC component \citep{Kush13,Kush18a,Kush18b}.
Thus, these simultaneous multi-wave observations help us to separate the Synchrotron and the IC emission components.

Figure \ref{fig:SED} shows the broadband UV to X-ray SEDs of the source for the three observations along with the contemporaneous data at NIR and optical bands from the
public archive of SMARTS \citep{Bon12} and Stewards \citep{Smith09} observatories as
well as published literature \citep{Hoso20}.  These SEDs clearly show strong  spectral evolution at both NIR-optical/FUV and the X-ray energies.
We have plotted three UV SEDs corresponding to a given SXT observation showing the maximum, minimum,
and the average SED corresponding to each observation to account for the likely presence of variability of a few percent pointed out in the previous section. The daily monitoring with {\it Swift}-XRT data \cyan{\citep{Komo20}} during the SXT observations do not show any appreciable changes in either the spectral state or flux \citep[see also][]{Komo21a}. Therefore, we derived a single integrated SXT spectrum from each year's {\it AstroSat} observation, and used them for the construction of the SEDs.

\begin{figure}
	\includegraphics[width=\columnwidth]{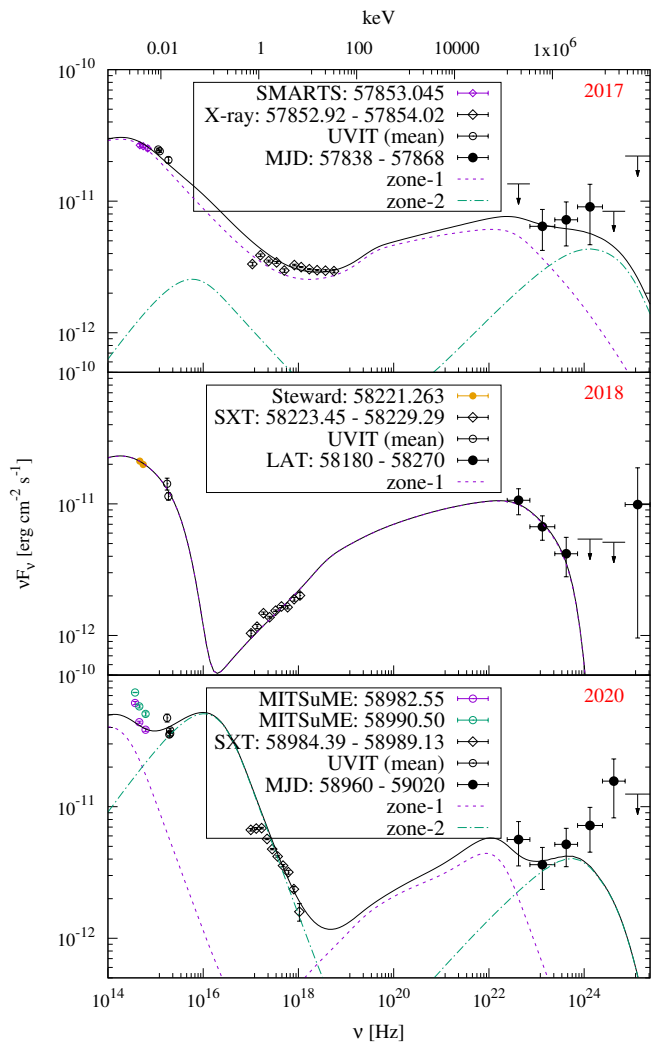}
    \caption{Modelled broadband SEDs constructed from the simultaneous/contemporaneous MW observations. The dashed and dotted-dashed curves show the broadband emission spectrum of the zone-1 and zone-2 respectively while the black curve is the sum total. For the 2018 broadband SED, a one zone model is sufficient to reproduce the overall MW emission (ref \S\ref{sec:model}).}
    \label{fig:mwSED}
\end{figure}

The SED constructed from the 2018 observation, shown in the middle panel of Fig. \ref{fig:SED} corresponds to the lowest X-ray flux state of the source and the corresponding averaged/mean optical-FUV spectrum shows a truncation or a sharp decline that can be modelled with a broken power-law with power-law indices of 1.2 at optical-UV and 4.3 for a two-point slope at FUV energies (\(F_\nu \propto \nu^{-\alpha} \)), respectively. The corresponding X-ray 
spectrum is the hardest of the three epochs, while the MeV-GeV spectrum (ref. Figure \ref{fig:mwSED}) is similar to the typical spectrum associated with the LBL state of the OJ 287 \citep{Abdo10,Kush13}. For the 2017 observation, the optical-FUV spectrum is relatively flatter and so are the X-ray and MeV-GeV spectra. The 2020 observation, on the other hand, has a harder optical-FUV spectrum but an extremely soft X-ray spectrum -- typical of HBL blazars when the synchrotron peak is around
UV energies \citep{Balo16}, while  the corresponding MeV-GeV spectrum shows a hardening above \(\sim\) 1 GeV \citep{Kush20b}. The spectral hardening at the high-energy end of the optical-UV and MeV-GeV spectra of the 2020 SED indicates that the emission is likely due the combination of the 
typical LBL spectral state of OJ~287 with the re-emergence of an HBL component \citep{Kush20b}.

 A closer look at the 2017 optical-UV spectrum reveals a comparatively flatter optical-UV and X-ray spectrum. Since the UVIT data do not show any sign of spectral cutoff as seen in 2018, it is possible that the flatness of
 the X-ray spectrum is due to the extension of the optical-UV spectrum to the X-ray energies  \citep[e.g.][]{Isobe01,Main20}. However, considering that the 2017 observation corresponds to the end-phase of
 the HBL component observed during a high X-ray activity period \citep{Kush18b}, the flatness could also be due to the weakening of the HBL component. 
In all, the three observations presented here,
suggest a spectral evolution of the source from the end-phase of activity driven by an additional HBL-like component in 2017 to its complete disappearance in 2018 and its revival in 2020. 
We have, therefore, modelled the 2018 observation with a one-zone leptonic model assuming a broken power-law particle distribution \citep[e.g.][]{Kush13}, where the optical-UV emission is due to synchrotron process, while the X-ray emission comes from Synchrotron Self-Compton (SSC) process, and the MeV-GeV $\gamma$-rays are due to external Comptonization \citep[EC;][]{Kush13,Arsioli18} of a \(\sim 250\) K thermal photon field. The 2017 and 2020 SEDs which require an additional HBL like emission component are modelled with a two-zone model with the additional zone also having a broken power-law particle distribution \citep[e.g.][]{Kush18b}).

\begin{table*}
\centering
\caption{Parameters and their values used for modelling the SEDs.}
\begin{tabular}{l c c c}
\hline
Parameters & 2017 & 2018 & 2020  \\
\hline
Particle index before break (p)  & 2.6 (2.0) & 2.5 & 2.0 (2.0) \\
Particle index after break (q) & 3.8 (3.8)  & 3.4 & 5.0 (5.0) \\
Magnetic field (Gauss) & 1.1 (0.34) & 0.7 & 1.5 (2.1) \\
Particle break energy $(\gamma_b^\ast)$ & 2280 (18536) & 2505  & 1364 (11817) \\
Relativistic particle energy density (erg cm$^{-3}$)& $\rm9.0\times10^{-3}$ ($\rm1.1\times10^{-2}$) & $\rm3.1\times10^{-2}$ & $\rm3.1\times10^{-3}$ ($\rm9.3\times10^{-4}$)\\ 
Doppler factor & 17.0 (6) & 16.5 & 18.6 (13.7) \\
Jet power (logscale, erg/s) & 46.1 (47.0) & 46.0 & 45.4 (46.5) \\
Minimum electron Lorentz factor$^\ast$ & 60 (60) & 20 & 45 (45) \\
Maximum electron Lorentz factor$^\ast$ & $1\times10^8$ ($1\times10^8$) & $1\times10^4$ & $1\times10^8$ ($1\times10^8$) \\
\hline
\multicolumn{3}{l}{Size of the emission region: $\rm 3 \times 10^{16}$ cm} \\
\multicolumn{3}{l}{Jet angle to the line of sight: $\rm 3^\circ$} \\
\multicolumn{3}{l}{$^\ast$in units of electron rest mass energy} \\
\multicolumn{3}{l}{Values in parentheses correspond to the second zone (see \S\ref{sec:model})} \\
\end{tabular}
\label{tab:parSEDs}
\end{table*}

In the 2-zone model, however, the number of plausible parameter sets and their values are many and the observed spectrum can be reproduced by varying the contribution from the two zones such that the sum remains at the observed level. An exhaustive study is beyond the scope of the present work and will be presented elsewhere. Here, we have tried a conservative approach keeping most of the parameters similar for both the zones.  For a given field of external seed photons, the HBL zone with its synchrotron peak at UV energies will automatically result in a second hump at higher energies. Thus, the hardening of MeV-GeV spectrum can be reproduced by either of the two mechanisms: SSC or EC. However, following the current status that most of the HBL/HSP blazars can be explained by the SSC, we have adopted the same approach here by estimating parameters in such a way that the
contribution of EC-IR from the 2nd-zone is negligible.
The SSC interpretation is consistent with the appearance of an additional SED peak around UV energies as well as the steepness of the X-ray spectrum that is
typically seen in HBLs \citep{Balo16,Yuan14} when the synchrotron peak is at UV energies \citep{Kush18b}.

A possible set of parameters for the three SEDs are given in Table \ref{tab:parSEDs} and the corresponding model SEDs are shown in figure \ref{fig:mwSED}. We find that to reproduce the flatter
X-ray spectrum of 2017, a weaker HBL component as well as the contribution of the optical-FUV synchrotron emission are required with the latter contributing dominantly.
Figure \ref{fig:mwSED} also shows that the MeV-GeV spectrum obtained in 2020 cannot be reproduced well. We would like to point out, however, that the optical, UV, and X-ray observations are for time-scales of a few hours to days, while the LAT MeV-GeV spectra are from 2 -- 3 months data as the source is too weak to be observed at shorter time-scales. Thus, LAT spectra represent an average state of the source instead of its spontaneous state seen in the optical to X-rays.

In the case of a single emission zone, Lorentz factor corresponding to the break in the particle spectrum, \(\gamma_b\), constrains the
location of the SED peak, and the observed spectra before and after the SED peak trace the particle
indices: (p) before,  and (q) after the peak (e.g., see \citep{Kush13}.  The optical-UV spectrum is, therefore, directly related to the particle spectral index, q. 
The same holds for the SSC component if the observed spectrum is away from the SED peak. Thus, for the 2018 observation one can estimate the particle indices p and q, from the optical-UV and X-ray spectrum respectively.
However, for the 2020 SED, the second zone appears dominant and thus, the X-ray spectrum constrains only the particle index after the SED peak viz., q. For the other index, we simply used the spectral
index from the modeling of the 2016--2017 activity \citep{Kush18b} that was also due to an  HBL-like component.  For the size of emission region, the jet angle to the line of sight, and the external photon field, we adopted a value used/derived from modelling the SEDs associated with the 2009 activity of the source \citep{Kush13}.

Considering all the above mentioned scenarios, we estimated the particle Lorentz factor corresponding to the SED peak (\(\gamma_b\)) by demanding that the low-energy-hump peak at NIR-optical energies (UV-soft X-ray for the second zone). However, since there is a degeneracy between the magnetic field and the Doppler factor ($\delta$) \citep{Kush13}, we started with a magnetic field value of unity and an appropriate $\delta$. With this assumption, the only unknown parameter left in the EC was the particle spectrum normalization which was derived by demanding that the model reproduce the MeV-GeV emission. A re-calibrated value for the magnetic field was then estimated by demanding that it reproduces the observed synchrotron component. With all the parameters thus derived, the SSC flux output from the model can be compared with the observed X-ray flux.  We used the method of trial and error for different $\delta$ values until we got the parameter set that reproduced the X-ray flux.  We, thus, estimated the parameter values for the  2018 SED for which the optical and the X-ray data provided the particle spectral indices
p and q respectively. For the 2020 SED, the second zone is dominant and the  X-ray spectrum constrains only the particle spectral index (q) after the break. Therefore, for the other index (p), we simply adopted a value used for modelling of the 2016--2017 HBL SEDs. The same values of the two indices were used for the first zone of emission as well. We then demanded that the modelled spectrum should reflect the spectral change observed in the MeV-GeV spectrum. 
For the 2017 SED, the optical-UV spectrum again  constrains only the particle index (q) after the SED peak of the first zone. Thus, for the index before the break (p), we adopted the value used for the intermediate state of the SED seen in 2009 \citep{Kush13}. For the second zone, we used the particle index before the SED peak from the model parameter of the 2020 SED, and adopted the values of q from the first zone (optical-FUV spectrum).
Note that $\gamma_{max}$ is unconstrained for 2017 and 2020 SEDs, while the spectral cutoff of 2018 constrain it to a value that is also supported by the MeV-GeV spectrum.  In the case of the 2017
SED, the reproduction of X-ray spectrum requires an extension of the synchrotron model to X-ray bands thus providing only a lower bound to $\rm\gamma_{max}$.  Furthermore, the excess apparent in the lowest optical data of the 2020 SED over a power-law form could be the 
accretion-disk component that was seen during the 2020 flare \citep{Kush20b}'.

\section{Discussion}
 OJ~287 has undergone spectrally distinct MW activity phases since mid-November 2015 \citep[e.g.][]{Gup17}. The first phase of the activity from end-2015 to mid-2016 showed a break in optical-NIR spectrum and a hardening in MeV-GeV \citep{Kush18a}. This was followed almost immediately by a phase that was driven by the presence of additional HBL component \citep[mid-2016 to mid-2017;][]{Kush18b} The source then gradually returned to its typical LBL spectral state in 2018, but only for a relatively short duration \citep[\(\sim 50\) days;][]{Kush20b}. The HBL component appears to have re-emerged subsequently, gradually becoming stronger, leading to its peak in April 2020 \citep{Komo20}. 
Our analysis of three {\it AstroSat} observations presented here, shows that the 2017 observation corresponds to the end-phase of the 2016--2017 HBL dominated activity, the 2018 observation to the disappearance of HBL dominated phase, 
while the 2020 observation caught the source just after its peak activity of the re-emerged HBL phase. Thus, our observations represent different moments of the evolutionary cycle of the HBL like emission component.

The strong and rapid MW variability \citep{Kap18,Komo20,Komo21a}, strong polarization \citep{Gup17,Gup19,Komo20,Komo21b}, and similarity of the broadband spectrum to HBL \citep{Kush20b,Kush18b} strongly suggest that the extremely soft X-ray state of OJ~287 originated in a relativistic jet. The broadband SED of 2020 presented here is simply a consequence of this additional strong HBL component appearing on top of the typical SED (LBL) of the source, as is clear from the lowest MeV-GeV data point as well as the hardening of the optical-UV spectrum. Overall, the broadband SED is well reproduced by a two-zone emission scenario (ref fig. \ref{fig:mwSED}). In this, the optical-UV to X-ray emission is due to synchrotron process and the MeV-GeV emission is due to external Comptonization and SSC processes (ref \S\ref{sec:model}) with an  HBL component being responsible for the hardening seen in the optical-FUV and the $\gamma$-ray spectra. The excess with respect to the model spectrum seen at the optical wavelengths could be due to emission from an accretion-disk, as was discovered during 2015--2016 activity \citep{Kush18a} and also  seen during the 2020 activity \citep{Kush20a}. The disagreement seen in the MeV-GeV emission is perhaps due to the $\gamma$-ray spectrum having been extracted from a much larger duration. The 2018 SED showing a cutoff/strong-steepening in the optical-FUV spectrum and the hardest of all X-ray spectrum is consistent with a one-zone emission scenario. 

The SED observed in 2017 is intermediate to that seen between 2020 and 2018 with an almost flat X-ray and MeV-GeV spectra (ref figs \ref{fig:SED} and \ref{fig:mwSED}). Of the two possibilities
mentioned earlier (ref. \S\ref{sec:model}) -- the continuation of synchrotron to X-ray energies or the HBL component for the flat X-ray spectrum, we found that the flattening is more likely due to the synchrotron component of the 1st-zone, with minimal contribution by the HBL-like component (2nd-zone). The 2nd-zone instead contributes significantly to the MeV-GeV spectrum via its SSC emission. Together with inferences from the 2020 SED, this suggests that X-ray spectral flatness in the LBL state is primarily due to the evolution of high-energy end of the synchrotron spectrum.

A spectral steepening/cutoff in the optical-UV spectrum during a low-flux state, similar to the one seen in the 2018 optical-UV spectrum can also be inferred for the low/weaker phase of the HBL driven MW activity during 2016--2017. 
The spectral evolution of this phase shows a flatter optical-UV and X-ray spectra. The X-ray emission is, however, almost an order of magnitude below the optical-UV level \citep[fig. 4, bottom panel plots]{Kush18b}, implying a strong spectral steepening/cutoff in the optical-UV spectrum. 
Further, the steepening/cutoff seems to happen around similar energies -- around FUV bands as is the case with 2018 spectrum. 
Since the observed spectrum is a direct reflection of underlying particle distribution in the case where most of the emission is from one emission region, these findings reveal that the high-energy-end of the particle spectrum is highly dynamic and the spectral steepening/cutoff strongly constrains the highest particle energies.  This is corroborated by the modelling of the  2017 and 2018 SEDs.  
For an assumed form of the particle distribution, the 2018 SED requires a much lower $\gamma_{max}$ while the 2017 optical-UV spectrum, provides only a lower bound in order to reproduce the observed X-ray spectrum (see Table \ref{tab:parSEDs}).  
This is consistent with our previous statement that almost all the observed X-ray spectral changes seen in the normal (LBL) spectral states are driven primarily by the evolution of the high-energy-end of the synchrotron component. Additionally, the detection of the source in a few of the lower energy bands of LAT during 2018 and the spectral cut-off revealed by the NUV-FUV data nicely demonstrates the one zone scenario that is usually invoked for modelling the broadband SED.

A comprehensive investigation of X-ray and MW variations of OJ 287 that encompasses the same  observation times as that of our \textit{AstroSat} observations has been presented recently in \citet{Komo21a}, \citet{Prince2021a}, and \citet{Prince2021b}. In particular, \citet{Komo21a} have employed a two-component scenario to explain the diversity of spectral changes seen in OJ 287. In the broadband context, the two-component scenario is similar to the two-zone emission invoked for modeling the broadband SEDs of the 2016--2017 activity \citep{Kush18b}. The broadband SED modelling presented in \citet{Prince2021b} focused mainly on the variations seen in soft X-ray flux and the associated MeV-GeV spectrum using the generally invoked synchrotron and SSC emission mechanisms. Our work here has additionally considered the normal LBL emission as revealed by the lowest-energy $\gamma$-ray data and the optical-UV spectrum.

OJ~287 has been argued to host a sub-parsec binary SMBH on the basis of quasi periodic optical outbursts (QPOOs) every $\sim 12$-yr \citep[and references therein but see  \citet{Br18,Butu20}]{Sill88,Dey18}. 
The latest of these outbursts occurred in 2015-end \citep{Valt16} and mid-2019 \citep{Laine20}.
The coincident appearance of spectral changes around these QPOOs -- in 2015--2016 \citep{Kush18a,Main20}, 2016 -- 2017 \citep{Komo17,Grupe17,Kush18b,Komo20,Komo21a}, and in
2020 \citep{Komo20,Komo21a,Komo21b,Kush20b} as well as in the earlier X-ray observations \citep[and references therein]{Isobe01} suggest a connection between the two. 
The implications of these spectral changes vis-a-vis the two classes of proposed models where one model attributes QPOOs to disk-impact in BBH scenario \citep{Dey18} while the
other to jet precession in either single or binary SMBH configuration \citep[e.g.][and references therein]{Br18,Butu20} are discussed in \citet{Kush20a}, \citet{Komo20}, and \citet{Komo21b}. 
In the jet precession scenario, only achromatic flux boosting is expected without any spectral changes and thus, the sharpness of the NIR-optical spectral break is inconsistent with broadband jet emission. This combined with the need for an additional HBL-like emission component currently favour the disk-impact BBH model \citep{Dey18}. The disk-impact BBH model also argues for a follow-up impact-induced jet activity but does not say anything about the spectral shape. 
Recently, \citet{Komo20} and \citet{Komo21b} attributed the soft X-ray spectrum dominated activity as the impact-induced jet activity. If true, then this implies that the impact-induced jet activity has a completely different broadband spectral shape (HBL-like) compared to the well-regarded spectral state (LBL) of the source. 
Alternatively, \citet{Huang2021}, argue that the soft X-ray state could due to a tidal event. In either scenario, the onset of the NIR-optical spectral break and the peculiarity of timing features reported during the soft X-ray states indicate some missing physics.

The extended period ($\sim$ 4 years) of activity representing very different SED phases is rarely seen among blazars. Even in a few cases where such changes have been reported, the duration is rather short (weeks to months) and limited to an on-going MW activity \citep[e.g.][]{Pian98,Hay15,Magic20}. The duration and strength of these changes in the SEDs imply drastic alterations of the physical conditions within the system. Overall, the findings reported here, and their implications involve all the issues ranging from accretion, jet-disk connection to relativistic particle energization, thus making OJ~287 an ideal laboratory for exploring a complete  spectrum of problems associated with accretion onto SMBH and relativistic jets. 
A continued higher cadence MW monitoring covering the next outburst predicted to be in 2022 \citep{Dey18} along with inputs from the EHT is expected to provide tighter constraints to all these issues and significantly enhance our current understanding.

\section{Summary and Conclusions}
The three observations of OJ~287 from 2017 to 2020 with {\it AstroSat} presented here  captured OJ~287 in different intensity and broadband spectral states.  The X-ray spectrum in the 0.3--7.0 keV energy band is found to be the hardest in 2018, while the simultaneous optical-FUV spectrum suggests a steep spectrum. The 2020 X-ray spectrum is the softest but the optical-FUV spectrum is the hardest. The X-ray and optical-FUV spectrum during 2017 is in-between these two states with a relatively flatter X-ray spectrum, consistent with that observed contemporaneously with the{\it NuSTAR}. The contemporaneous MeV-GeV emission from LAT shows that the three SEDs trace the  evolution of the source from the end-phase of an activity driven by an additional HBL-like emission component in 2017 to its disappearance in 2018, followed by its revival in 2020.

The broadband SED associated with the 2018 {\it AstroSat} observation is similar to the
normal LBL state of OJ 287 and a single zone leptonic scenario with synchrotron, and SSC or EC-IR processes can reproduce the broadband SED. The SED from the 2020 observations,
shows a hardening at the high-energy end of the optical-FUV and MeV-GeV spectra with respect to the LBL state of OJ 287. Such a hardening is inconsistent with the normal one-zone scenario. 
Instead, the optical-FUV emission together with the X-ray spectrum indicates an additional HBL-like emission component and the 2020 SED is well-reproduced in a two-zone emission
scenario with the additional zone emitting an HBL spectrum. The SED observed in 2017 is intermediate to that reported for 2018 and 2020, and corresponds to the weakening or end of the phase driven by an HBL-like component that occurred in 2016--2017 (thus resembling the activity seen in 2020). 
The spectral profile of the optical-FUV spectrum and the flat MeV-GeV spectrum are not reproduced in single-zone emission scenario and need an additional HBL-like emission component.
A two-zone model, similar to that used for the 2020 observation, but with a much weaker HBL component well-reproduces the broadband SED. The modelling shows that the nearly flat X-ray spectrum is primarily due to the continuation of the synchrotron spectrum to X-ray energies from the 1st-zone (LBL) while the flat MeV-GeV spectrum is primarily due to SSC component from the 2nd-zone (HBL).

The availability of the FUV data from {\it AstroSat} provides a direct view of the evolution of
the high-energy-end of the optical-UV spectrum and leads to strong constraints on the contribution
of a synchrotron component (primarily $\gamma_{max}$) in driving the observed X-ray spectral changes. The modelling of the 2017 and 2018 optical SED, and constraints on the spectral shape of the new component from the 2020 SED suggest that in the normal LBL spectral state of 
OJ~287, the X-ray spectral changes are primarily driven by synchrotron emission in the optical-UV extending to X-rays energies.

\section*{Acknowledgements}
The authors thank the anonymous referee for constructive inputs that have helped in improving the overall presentation. PK acknowledges ARIES Aryabhatta Postdoctoral Fellowship (A-PDF) grant (AO/A-PDF/770). M.P. thanks the ﬁnancial support of UGC, India program through DSKPDF fellowship (grant No. BSR/
2017-2018/PH/0111). We thank the Indian Space Research Organisation for scheduling the observations within a short period of time and the Indian Space Science Data Centre (ISSDC) for making the data available.
The 2017 and 2018 data were acquired as part of joint effort with the Event Horizon Telescope collaboration and we gratefully acknowledge their permission to present these data here.  The 2020 data (PI: A. Agarwal) were acquired under the target of opportunity program of ISRO.
This work has been performed utilizing the calibration data-bases and auxillary analysis tools developed, maintained and distributed by {\it AstroSat}-SXT team with members from various institutions in India and abroad and the  SXT Payload Operation Center (POC) at the TIFR, Mumbai for the pipeline reduction.  The work has also made use of software, and/or web tools obtained from NASA's High Energy Astrophysics Science Archive Research Center (HEASARC), a service of the Goddard Space Flight Center and the Smithsonian Astrophysical Observatory. This paper has made use of up-to-date SMARTS optical/near-infrared light curves that are available at \url{www.astro.yale.edu/smarts/glast/home.php}.
Data from the Steward Observatory spectro-polarimetric monitoring project were used. This
program\footnote{\url{http://james.as.arizona.edu/~psmith/Fermi/}} is supported by Fermi Guest Investigator grants NNX08AW56G, NNX09AU10G, NNX12AO93G, and NNX15AU81G.\\

\section*{Facilities:} $AstroSat$, $NuSTAR$, $Fermi$, SMARTS, Steward\\
\section*{Data Availability:}
{\it AstroSat} data for all the observations used in this paper are publicly available from the {\it AstroSat} archives maintained by the ISSDC, Bengaluru: $https://astrobrowse.issdc.gov.in/astro\_archive/archive/Home.jsp$.  The {\it NuSTAR} data are publicly available from the archives of HEASARC maintained by NASA. SMARTS and Steward data are available
publicly at their respective archives.\\

\bibliographystyle{mnras}
\bibliography{example} 

\begin{thebibliography}{99}
\bibitem[\protect\citeauthoryear{Abdo et al.}{2010}]{Abdo10} Abdo A.~A., Ackermann M., Agudo I., Ajello M., Aller H.~D., Aller M.~F., Angelakis E., et al., 2010, ApJ, 716, 30. doi:10.1088/0004-637X/716/1/30

\bibitem[\protect\citeauthoryear{Abdollahi et al.}{2020}]{4fgl} Abdollahi S., Acero F., Ackermann M., Ajello M., Atwood W.~B., Axelsson M., Baldini L., et al., 2020, ApJS, 247, 33. doi:10.3847/1538-4365/ab6bcb

\bibitem[\protect\citeauthoryear{Antia et al.}{2017}]{Ant17}
Antia, H. M., Yadav, J. S., Agrawal, P. C., et al. 2017, ApJS, 231, 10

\bibitem[\protect\citeauthoryear{Arnaud}{1985}]{Ar1985}
Arnaud, K.A., Branduardi-Raymont, G., Culhane, J. L., Fabian, A. C., Hazard, C., McGlynn, T. A., Shafer, R. A., Tennant, A. F., Ward, M. J.et al. 1985, MNRAS, 217, 105.

\bibitem[\protect\citeauthoryear{Arnaud}{1996}]{Ar1996}
Arnaud, K.A., 1996, Astronomical Data Analysis Software and Systems V, eds. G. Jacoby and J. Barnes,p17, ASP Conf. Series volume 101.

\bibitem[\protect\citeauthoryear{Arsioli \& Chang}{2018}]{Arsioli18} Arsioli B., Chang Y.-L., 2018, A\&A, 616, A63. doi:10.1051/0004-6361/201833005

\bibitem[\protect\citeauthoryear{Asplund et al}{2009}]{As2009}
Asplund M., Grevesse N., Sauval A.J. \& Scott P. 2009, ARAA, 47, 481

\bibitem[\protect\citeauthoryear{Atwood et al.}{2009}]{lat09} Atwood W.~B., Abdo A.~A., Ackermann M., Althouse W., Anderson B., Axelsson M., Baldini L., et al., 2009, ApJ, 697, 1071. doi:10.1088/0004-637X/697/2/1071

\bibitem[\protect\citeauthoryear{{Bekhti} et~al./}{{Bekhti}
 et~al.}{2016}]{Bekhti2016} {Ben~Bekhti} N., et~al.,2016, A\&A, 594, A116.

\bibitem[\protect\citeauthoryear{Balokovi{\'c} et al.}{2016}]{Balo16} Balokovi{\'c} M., Paneque D., Madejski G., Furniss A., Chiang J., Ajello M., Alexander D.~M., et al., 2016, ApJ, 819, 156. doi:10.3847/0004-637X/819/2/156

\bibitem[\protect\citeauthoryear{Bonning et al.}{2012}]{Bon12} Bonning E., Urry C.~M., Bailyn C., Buxton M., Chatterjee R., Coppi P., Fossati G., et al., 2012, ApJ, 756, 13. doi:10.1088/0004-637X/756/1/13

\bibitem[\protect\citeauthoryear{Brien \& VERITAS Collaboration}{2017}]{Brien17} Brien S.~O., VERITAS Collaboration, 2017, ICRC, 301, 650.

\bibitem[\protect\citeauthoryear{Britzen et al.}{2018}]{Br18}
Britzen, S., Fendt, C., Witzel, G., Qian, S. -J., Pashchenko, I.~N., Kurtanidze, O, Zajacek, M., Martinez, G., Karas, V., Aller, M., Aller, H., Eckart, A., Nilsson, K., Ar{\'e}valo, P., Cuadra, J., Subroweit, M., Witzel, A., 2018, MNRAS, 478, 3199B.

\bibitem[\protect\citeauthoryear{Butuzova \& Pushkarev}{2020}]{Butu20} Butuzova M.~S., Pushkarev A.~B., 2020, Univ, 6, 191. doi:10.3390/universe6110191

\bibitem[\protect\citeauthoryear{Cardelli, Clayton \& Mathis}{1989}]{Cardelli89} Cardelli J.~A., Clayton G.~C., Mathis J.~S., 1989, ApJ, 345, 245. doi:10.1086/167900

\bibitem[\protect\citeauthoryear{Cohen}{2017}]{Co17}
Cohen, M., 2017, Galax, 5, 12C.

\bibitem[\protect\citeauthoryear{Dey et al.}{2021}]{Dey21} Dey L., Valtonen M.~J., Gopakumar A., Lico R., G{\'o}mez J.~L., Susobhanan A., Komossa S., et al., 2021, MNRAS.tmp. doi:10.1093/mnras/stab730

\bibitem[\protect\citeauthoryear{Dey et al.}{2018}]{Dey18}
Dey, L., Valtonen, M.~J., Gopakumar, A., Zola, S., Hudec, R., Pihajoki, P., Ciprini, S. et al., 2018, ApJ, 866, 11D.

\bibitem[\protect\citeauthoryear{Fossati et al.}{1998}]{Fo1998} 
Fossati G., Maraschi L., Celotti A., Comastri A., Ghisellini G., 1998, MNRAS, 299, 433. doi:10.1046/j.1365-8711.1998.01828.x

\bibitem[\protect\citeauthoryear{Grupe et al.}{2017}]{Grupe17} Grupe D., Komossa S., amp, Falcone A., 2017, ATel, 10043

\bibitem[\protect\citeauthoryear{Gupta et al.}{2017}]{Gup17}
Gupta A.~C., Agarwal A., Mishra A., Gaur H., Wiita P.~J., Gu M.~F., Kurtanidze O.~M., et al., 2017, MNRAS, 465, 4423. doi:10.1093/mnras/stw3045

\bibitem[\protect\citeauthoryear{Gupta et al.}{2019}]{Gup19}
Gupta A.~C., Gaur H., Wiita P.~J., Pandey A., Kushwaha P., Hu S.~M., Kurtanidze O.~M., et al., 2019, AJ, 157, 95. doi:10.3847/1538-3881/aafe7d

\bibitem[\protect\citeauthoryear{Harrison et al.}{2013}]{Har13}
Harrison, F.A.; Craig, W.W.; Christensen, F.E. et al., 2013, ApJ, 770, 103.

\bibitem[\protect\citeauthoryear{Hayashida et al.}{2015}]{Hay15}
Hayashida, M., Nalewajko, K., Madejski, G.~M. et al., 2015, ApJ, 807, 79H.

\bibitem[\protect\citeauthoryear{Hosokawa et al.}{2020}]{Hoso20} Hosokawa R., Adachi R., Murata K.~L., Niwano M., Ogawa F., Nakamura N., Yatsu Y., et al., 2020, ATel, 13755

\bibitem[\protect\citeauthoryear{Huang et al.}{2021}]{Huang2021} Huang S., Hu S., Yin H., Chen X., Alexeeva S., Gao D., Jiang Y., 2021, arXiv, arXiv:2106.14368

\bibitem[\protect\citeauthoryear{Isobe et al.}{2001}]{Isobe01} Isobe N., Tashiro M., Sugiho M., Makishima K., 2001, PASJ, 53, 79. doi:10.1093/pasj/53.1.79


\bibitem[\protect\citeauthoryear{Joseph et al.}{2021}]{Joseph21} Joseph P., Stalin C.~S., Tandon S.~N., Ghosh S.~K., 2021, arXiv, arXiv:2101.06377

\bibitem[\protect\citeauthoryear{Kapanadze et al.}{2018}]{Kap18}
Kapanadze, B., Vercellone, S., Romano, P., Hughes, P., Aller, M., Aller, H., Kapanadze, S., Tabagari, L., 2018, MNRAS, 580, 407K

\bibitem[\protect\citeauthoryear{Komossa et al.}{2021b}]{Komo21b} Komossa S., Ciprini S., Dey L., Gallo L.~C., Gomez J.~L., Gonzalez A., Grupe D., et al., 2021, arXiv, arXiv:2104.12901

\bibitem[\protect\citeauthoryear{Komossa et al.}{2017}]{Komo17} Komossa S., Grupe D., Schartel N., Gallo L., Gomez J.~L., Kollatschny W., Kriss G., et al., 2017, IAUS, 324, 168. doi:10.1017/S1743921317001648

\bibitem[\protect\citeauthoryear{Komossa et al.}{2021a}]{Komo21a} Komossa S., Grupe D., Parker M.~L., G{\'o}mez J.~L., Valtonen M.~J., Nowak M.~A., Jorstad S.~G., et al., 2021, MNRAS, 504, 5575. doi:10.1093/mnras/stab1223

\bibitem[\protect\citeauthoryear{Komossa et al.}{2020}]{Komo20} Komossa S., Grupe D., Parker M.~L., Valtonen M.~J., G{\'o}mez J.~L., Gopakumar A., Dey L., 2020, MNRAS, 498, L35

\bibitem[\protect\citeauthoryear{Kushwaha et al.}{2016}]{Kush16}
Kushwaha, P., Chandra, S., Misra, R., Sahayanathan, S., Singh, K. P. \& Baliyan, K. S.,2016, ApJ, 822, 13K.

\bibitem[\protect\citeauthoryear{Kushwaha, Sahayanathan \& Singh}{2013}]{Kush13}
Kushwaha P., Sahayanathan S., Singh K.~P., 2013, MNRAS, 433, 2380 

\bibitem[\protect\citeauthoryear{Kushwaha et al.}{2017}]{Kush17}
Kushwaha, P., Sinha, A., Misra, R. and Singh, K. P.,  de Gouveia Dal Pino, E. M. 2017, ApJ, 849, 138

\bibitem[\protect\citeauthoryear{Kushwaha et al.}{2021}]{KushICRC2021} Kushwaha P., Singh K.~P., Sinha A., Pal M., Dewangan G., Agarwal A., 2021, PoS(ICRC2021), 395, 644

\bibitem[\protect\citeauthoryear{Kushwaha et al.}{2018a}]{Kush18a}
Kushwaha, P., Gupta, A.~C., Wiita, P.~J., Gaur, H., de Gouveia Dal Pino, E.~M., Bhagwan, J., Kurtanidze, O.~M., Larionov, V.~M., Damljanovic, G. et al. 2018, MNRAS, 473, 1145K.

\bibitem[\protect\citeauthoryear{Kushwaha et al.}{2018b}]{Kush18b}
Kushwaha, P., Gupta, A.~C., Wiita, P.~J., Pal, M., Gaur, H., de Gouveia Dal Pino, E.~M., Kurtanidze, O.~M., et al, 2018, MNRAS, 479, 1672K.

\bibitem[\protect\citeauthoryear{Kushwaha et al.}{2020}]{Kush20b} Kushwaha P., Pal M., Kalita N., Kumari N., Naik S., Gupta A.~C., de Gouveia Dal Pino E.~M., et al., 2020b, arXiv, arXiv:2010.14431

\bibitem[\protect\citeauthoryear{Kushwaha}{2020}]{Kush20a} Kushwaha, P., 2020a, Galex, 8, 15K.

\bibitem[\protect\citeauthoryear{Laine et al.}{2020}]{Laine20} Laine S., Dey L., Valtonen M., Gopakumar A., Zola S., Komossa S., Kidger M., et al., 2020, ApJL, 894, L1. doi:10.3847/2041-8213/ab79a4

\bibitem[\protect\citeauthoryear{MAGIC Collaboration et al.}{2020}]{Magic20} MAGIC Collaboration, Acciari V.~A., Ansoldi S., Antonelli L.~A., Babi{\'c} A., Banerjee B., Barres de Almeida U., et al., 2020, A\&A, 637, A86. doi:10.1051/0004-6361/201834603

\bibitem[\protect\citeauthoryear{Main Pal et al.}{2020}]{Main20}
Main Pal, Kushwaha, P., Dewangan, G. C. \& Pawar, P. K. 2020, ApJ, 890, 47.

\bibitem[\protect\citeauthoryear{Massaro et al.}{2004}]{Mass2004}
Massaro, E., Perri, M., Giommi, P. \& Nesci, R., 2004, A\&A, 413, 489.

\bibitem[\protect\citeauthoryear{Nilsson et al.}{2010}]{Nil10}
Nilsson, K., Takalo, L.~O., Lehto, H.~J. \& Sillanp{\"a}{\"a}, A., 2010, A\&A, 516A, 60N.

\bibitem[\protect\citeauthoryear{Pian et al.}{1998}]{Pian98}
Pian, E., Vacanti, G.,  Tagliaferri, G., Ghisellini, G., Maraschi, L., Treves, A., Urry, C. M., Fiore, F. Giommi, P., Palazzi, E., Chiappetti, L. Sambruna, R. M., 1998, ApJ, 492L, 17P

\bibitem[\protect\citeauthoryear{Prince et al.}{2021a}]{Prince2021a} Prince R., Raman G., Khatoon R., Agarwal A., Varun, Gupta N., Czerny B., et al., 2021, MNRAS, 508, 315. doi:10.1093/mnras/stab2545


\bibitem[\protect\citeauthoryear{Prince et al.}{2021b}]{Prince2021b}Prince R., Agarwal A., Gupta N., Majumdar P., Czerny B., Cellone S.~A., Andruchow I., 2021, A\&A, 654, A38. doi:10.1051/0004-6361/202140708


\bibitem[\protect\citeauthoryear{Rodr{\'\i}guez-Ram{\'\i}rez et al.}{2020}]{Rod20} Rodr{\'\i}guez-Ram{\'\i}rez J.~C., Kushwaha P., de Gouveia Dal Pino E.~M., Santos-Lima R., 2020, MNRAS, 498, 5424. doi:10.1093/mnras/staa2664

\bibitem[\protect\citeauthoryear{Siejkowski \& Wierzcholska}{2017}]{Siej17} Siejkowski H., Wierzcholska A., 2017, MNRAS, 468, 426. doi:10.1093/mnras/stx495

\bibitem[\protect\citeauthoryear{Sillanpaa et al.}{1996}]{Sill96a}
Sillanpaa, A, Takalo, L.~O., Pursimo, T., Lehto, H.~J., Nilsson, K., Teerikorpi, P, Heinaemaeki, P. et al. 1996 A\&A, 305L, 17S.

\bibitem[\protect\citeauthoryear{Sillanpaa et al.}{1988}]{Sill88} Sillanpaa A., Haarala S., Valtonen M.~J., Sundelius B., Byrd G.~G., 1988, ApJ, 325, 628. doi:10.1086/166033

\bibitem[\protect\citeauthoryear{Singh et al.}{1985}]{Si1985}
Singh, K. P., Garmire, G. P., Nousek, J. A., 1985, ApJ, 297, 633.

\bibitem[\protect\citeauthoryear{Singh et al.}{2014}]{Si2014}
Singh, K. P., Tandon, S. N., Agrawal, P. C., et al. 2014, Proc. SPIE, Space Telescopes and Instrumentation 2014: Ultraviolet to Gamma Ray. 9144, 91441S
doi:10.1117/12.2062667

\bibitem[\protect\citeauthoryear{Singh et al.}{2016}]{Si2016}
Singh K. P., Stewart, G. C., Chandra, S. et al., 2016, Proc. SPIE, in Space Telescopes and Instrumentation 2016: Ultraviolet to Gamma Ray. 9905, p. 99051E, doi:10.1117/12.2235309

\bibitem[\protect\citeauthoryear{Singh et al.}{2017}]{Si2017}
Singh, K. P., Stewart, G. C., Westergaard, N. J., et al. 2017, JApA, 38, 29.

\bibitem[\protect\citeauthoryear{Sitko and Junkkarinen}{1985}]{Sit85}
Sitko, M.~L. and Junkkarinen, V.~T., 1985, PASP, 97, 1158S.

\bibitem[\protect\citeauthoryear{Smith et al.}{2009}]{Smith09} Smith P.~S., Montiel E., Rightley S., Turner J., Schmidt G.~D., Jannuzi B.~T., 2009, arXiv, arXiv:0912.3621

\bibitem[\protect\citeauthoryear{{Tandon} et~al.,}{{Tandon} et~al.}{2017}]{Tan17}
{Tandon} S.~N.,  et~al., 2017, AJ, 154, 128.

\bibitem[\protect\citeauthoryear{{Tandon} et~al.,}{{Tandon} et~al.}{2020}]{Tan20}
{Tandon} S.~N.,  et~al., 2020, AJ, 2020, 159, 158.

\bibitem[\protect\citeauthoryear{Vaughan et al.}{2003}]{vaughan2003} Vaughan S., Edelson R., Warwick R.~S., Uttley P., 2003, MNRAS, 345, 1271. doi:10.1046/j.1365-2966.2003.07042.x

\bibitem[\protect\citeauthoryear{Valtonen et al.}{2010}]{Valt10} Valtonen M.~J., Mikkola S., Merritt D., Gopakumar A., Lehto H.~J., Hyv{\"o}nen T., Rampadarath H., et al., 2010, ApJ, 709, 725. doi:10.1088/0004-637X/709/2/725

\bibitem[\protect\citeauthoryear{Valtonen et al.}{2016}]{Valt16} Valtonen M.~J., Zola S., Ciprini S., Gopakumar A., Matsumoto K., Sadakane K., Kidger M., et al., 2016, ApJL, 819, L37. doi:10.3847/2041-8205/819/2/L37

\bibitem[\protect\citeauthoryear{Wenzel}{1971}]{Wenz71}
{Wenzel} W., 1971, Information Bulletin on Variable Stars, 593, 1W.

\bibitem[\protect\citeauthoryear{Yuan \& Fan}{2014}]{Yuan14} Yuan Y., Fan J., 2014, Ap\&SS, 352, 207. doi:10.1007/s10509-014-1878-y

\end{thebibliography}










\bsp	
\label{lastpage}
\end{document}